\documentclass[12pt]{iopart}

\usepackage{graphicx}
\usepackage{epstopdf}
\graphicspath{{./Figs/}} 

\usepackage{latexsym}

\expandafter\let\csname equation*\endcsname\relax
\expandafter\let\csname endequation*\endcsname\relax
\usepackage{amsmath}

\usepackage{amssymb}
\usepackage{bm}
\usepackage{wasysym}

\usepackage{ifthen}
\usepackage{color}
\usepackage[colorlinks,allcolors=blue,bookmarks=true]{hyperref}

\newcommand{\const}{\mbox{const}}

\newcommand{\braket}[1]{\left\langle #1 \right\rangle }
\newcommand{\avg}[1]{\left\langle #1 \right\rangle }

\newcommand{\var}{\mbox{\text{Var}}}

\newcommand{\beq}{\begin{eqnarray}}
\newcommand{\eeq}{\end{eqnarray}}

\newcommand{\hide}[1]{}  
\newcommand{\rmrk}[1]{{#1}} 
\newcommand{\rmrkR}[1]{{#1}} 

\newcommand{\Eq}[1]{{\textcolor{blue}{Eq.}}~\!\!(\ref{#1})} 
\newcommand{\Sec}[1]{{\textcolor{blue}{Sec.}}\!\!~(\ref{#1})} 
\newcommand{\App}[1]{\ref{#1}}  
\newcommand{\Fig}[1] {{\textcolor{blue}{Fig.}}~\!\!\ref{#1}}
\newcommand{\sect}[1]{{\bf #1.-- }}

\newcommand{\hrefl}[2]{\href{#2}{#1}}

\expandafter\ifx\csname url\endcsname\relax \def\href{#1}{[link]} \fi

\begin{document}

\title[Emergence of Sinai Physics]{Emergence of Sinai Physics in the stochastic motion of passive and active particles} 

\author{Dekel Shapira and Doron Cohen} 

\address{
\mbox{Department of Physics, Ben-Gurion University of the Negev, Beer-Sheva 84105, Israel} 
}

\begin{abstract}
A particle that is immersed in a uniform temperature bath performs Brownian diffusion, as discussed by Einstein. But Sinai has realized that in a ``random environment" the diffusion is suppressed. Follow-up works have pointed out that in the presence of bias~$f$ there are delocalization and sliding transitions, with threshold value $f_c$ that depends on the disorder strength. We discuss in a critical way the emergence of Sinai physics for both passive and active Brownian particles. Tight-binding and Fokker-Planck versions of the model are addressed on equal footing. We assume that the transition rates between sites are enhanced either due to a driving mechanism or due to self-propulsion mechanism that are induced by an irradiation source. Consequently, counter intuitively, the dynamics becomes sub-diffusive and the relaxation modes become over-damped. For a finite system, spontaneous delocalization may arise, due to residual bias that is induced by the irradiation.
\end{abstract}

\maketitle

\section{Introduction}

The most familiar type of stochastic motion is {\em random walk} 
along a tight binding chain (lattice sites are labelled by $n$), 
where all the transition rates are identical. 
The dynamics is described by a rate equation 
%
\beq\label{eq:rate-eq}
\dot{\bm{p}} \ = \ \bm{W} \bm{p}
\eeq
where ${\bm{p} = \{ p_n \} }$ are the probabilities to find the particle in any of the sites, 
and $\bm{W}$ is a tridiagonal matrix whose off-diagonal elements are the transition rates.   
In the continuum limit the dynamics is described by a diffusion equation.

The question arises whether normal diffusion is affected by disorder.   
The simplest type of disorder arises if the rates $w_n$ on each bond are random, 
aka bond-disorder or resistor-network-disorder.
Hereafter we assume that the bond-disorder is weak, 
such that the probability to observe a disconnected chain is zero. 
Accordingly, for such type of disorder, the dynamics remains `normal'.  

\sect{Sinai model}
Sinai had proposed to study a more general model 
that has been termed {\em random walk in random environment} \cite{Sinai1983}.
In this model the rates $w_n^{\pm}$ for forward and backward transitions 
over the $n$-th bond are uncorrelated (instead of being equal).
The stochastic field over the $n$-th bond is defined thorough 
the ratio, namely, ${ w_n^{+}/w_n^{-} = \exp[\mathcal{E}_n] }$.
Assuming that the $\mathcal{E}_n$ are uncorrelated, 
it follows that in the absence of bias (${\braket{\mathcal{E}}=0}$)    
the spatial spreading is {\em sub-diffusive}  ($x\sim\ln^2(t)$) 
rather than diffusive ($x\sim\sqrt{t}$).

\sect{Sliding transition}
Subsequent works by Derrida and followers \cite{Derrida1983,BOUCHAUD1990,BOUCHAUD1990a}
have considered the effect of non zero bias ${f \equiv \braket{\mathcal{E}} > 0 }$.   
The existence of a critical bias $f_c$ has been highlighted, 
such that  for ${f<f_c}$ the drift velocity vanishes, while for ${f>f_c}$ it becomes finite. 
This has been termed {\em sliding transition}.              
We note that Sinai has assumed that the $\mathcal{E}_n$ are strictly uncorrelated. 
The implications of long range correlations have been studied as well in \cite{Creep}, 
highlighting a crossover between Ohmic, Creep, and Sinai regimes. 
It is only in the latter case (of effectively uncorrelated disorder), 
that one observes a sharp sliding transition.

\sect{Delocalization transition}
From a completely different perspective Hatano, Nelson and followers~\cite{Hatano1996,Hatano1997,Shnerb1998,Feinberg1999,KAFRI2004,KAFRI2005}
have considered the {\em delocalization transition} for non-Hermitian Hamiltonians.
In the present context the role of the non-Hermitian Hamiltonian 
is taken by the real non-symmetric (and hence non-Hermitian) matrix $\bm{W}$  
that conserves probability (${\sum p_n =1}$).    
The details of the delocalization scenario in the stochastic context 
have been discussed in \cite{HurowitzCohen2016,HurowitzCohen2016a}. 
Here we are talking about the relaxation modes \rmrk{$\psi^{(r)}$} that are associated with eigenvalues $\lambda_r$ 
of a finite length ring. They are determined by the equation 
%
\beq
\bm{W} \psi \ = \ -\lambda \psi 
\eeq
The trivial eigenvalue ${\lambda_0=0}$ corresponds to the steady state. 
Delocalization is indicated by complex eigenvalues. 
For large enough $f$ the eigenvalues at the vicinity of ${\lambda=0}$ become complex, 
which implies under-damped relaxation.
The term `delocalization' reflects the insight that for a ring geometry 
under-damped relaxation towards the steady-state is associated 
with complex-valued relaxation-modes that are extended over the ring.  
It has been emphasized in \cite{HurowitzCohen2016} that the threshold field for this {\em finite-size-system} transition 
is smaller than the $f_c$ that is estimated for sliding along an infinite chain.

\subsection{The Physics behind Sinai Physics}  

The model of Sinai is rather artificial. 
One wonders what are the physical assumptions  
that imply the emergence of the sliding and the delocalization transitions. 
The stochastic filed is defined via ${ w_n^{+}/w_n^{-} = \exp[\mathcal{E}_n] }$. 
For a chain that is coupled to a single bath 
that has temperature $T$ the stochastic field is   
%
\beq \label{eBath}
\mathcal{E}_n[\text{single bath, no driving}] \ = \  \frac{V_n-V_{n{+}1}}{T}
\eeq
where $V_n$ is the on-site potential. 
This field exhibits telescopic correlations, 
that reflect the bounded energy landscape of $V_n$, 
while Sinai model assumes uncorrelated stochastic field  
whose potential features activation barriers 
of height $\propto \sqrt{L}$ for segments of length~$L$.

\begin{figure}
\centering
\includegraphics[width=9cm]{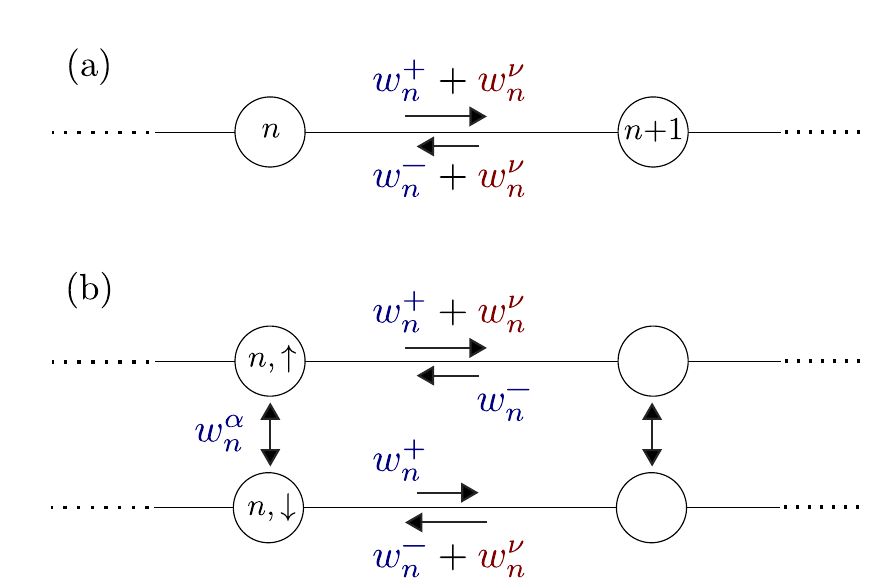}
\caption{ 
\textbf{(a)} Passive particle on a chain, sites labelled by~$n$. 
The transition rates $w_n^{\pm}$ are induced by a bath that has temperature~$T$.  
Extra transitions $w_n^{\nu}$ are induced by a driving source.
\textbf{(b)} Active particle on a chain. Such particle has an extra degree of freedom ${s=\uparrow,\downarrow}$ 
that indicates its orientation. The bath induced rates are $w_n^{\pm}$ in the $n$ coordinate. 
The flipping rate in the $s$-coordinate is $w_n^{\alpha}$. 
Extra transitions $w_n^{\nu}$ are induced by the propulsion.      
\label{f1}}
\end{figure}

\rmrkR{The emergence of Sinai Physics necessitates 
a mechanism that breaks the telescopic correlations of the stochastic field.} 
One simple option is to assume non-uniform temperature such that  ${ \mathcal{E}_n = [V_n-V_{n{+}1}]/T_n }$, 
with uncorrelated $T_n$ on each bond. 
In the present work we would like to consider 
more interesting setups that are illustrated in \Fig{f1}. 
\rmrkR{In both setups the system is a tight binding chain immersed 
in a bath that has a uniform temperature. 
In the absence of ``irradiation" the particle performs a diffusive motion.
With ``irradiation" the transition rates between sites are enhanced. 
The counter-intuitive expectation, based on Sinai's argument,
is to witness sub-diffusion instead of diffusion. 
To get an insight for this puzzling expectation, note that on short times 
one is likely to witness an enhanced spreading rate. 
But we focus on long times. 
Then it is important to notice that it is not 
the {\em rates} of transitions but the {\em topography} of the stochastic potential that matters 
(for argumentation purpose note that this topography is not affected 
if all the rates are multiplied by the same large factor). 
Due to the irradiation, the topography of the stochastic potential features an ever growing activation barriers 
that suppress the long time spreading, which is the essence of Sinai Physics.}

\subsection{\rmrkR{Paradigms for anomalous transport}}  

There are of course different paradigms for anomalous transport. 
But we focus here on what we call ``Sinai Physics".
For sake of clarity, following Sinai, we consider {\em weak} disorder: 
not only that the probability to observe a disconnected bond is zero, 
but furthermore it is assumed (for sake of argumentation) that all the rates are comparable. 
In particular it is implied that ${\mathcal{E}_n \ll 1}$. This weak disorder assumption 
allows in a later stage to adopt a very convenient continuum-limit Fokker-Planck approximation for the dynamics.

\rmrkR{The model of Sinai is strictly one-dimensional (1D). 
Merely from a geometrical point of view, local barriers can block transport. 
In more than 1D one generically encounters rugged landscape 
that features ponds and outlets, see e.g. discussion in \cite{stein1995broken}. 
Note however that signatures of Sinai Physics may arise 
for the dynamics along the infinite cluster 
in the vicinity of a percolation threshold.}   

\rmrkR{It is important to appreciate that ``Sinai Physics" is not the only mechanism for anomalous transport. Increasing the activity in conjunction with quenched disorder can lead to counter-intuitive results that are not necessarily related to Sinai Physics. Particle-particle interactions in combination with quenched disorder generically produce such effects. Ref.\cite{reichhardt2014active} has considered the run-and-tumble dynamics of disks that move in a
disordered array of fixed obstacles. Such system exhibits non-monotonic transport response due to crowding clogging and jamming; the motion is almost blocked and proceed in the form of avalanches \cite{reichhardt2018avalanche}; and under bias might exhibit negative differential mobility \cite{reichhardt2017negative}. This type of dynamics can be demonstrated also for a lattice version \cite{reichhardt2021clogging,bertrand2018optimized}.}

\subsection{Scope of this paper}

We consider two models for which Sinai Physics is likely to emerge. 
The models are illustrated in \Fig{f1}.  
In the absence of ``irradiation" the particle performs 
transitions that are dictated by \Eq{eBath}. 
In the presence of ``irradiation" the rates are enhanced.
Specifically:  
\begin{itemize}
\item {\em Passive-particle model.-- }
In the model that is described by \Fig{f1}a, 
the chain is irradiated by a high-temperature source.
The term `irradiation' does not have to be taken literally, 
it is just a convenient way to describe the effect of a second bath, 
that is regarded as a driving-source.    
For an infinite-temperature source the transition rates become ${ w_n^{\pm} + w_n^{\nu} }$. 
The extra rates ${ w_n^{\nu} }$ that are induced by the irradiation 
are not biased to any direction.   
\item {\em Active-particle model.-- }
In the model that is described by \Fig{f1}b, the irradiation induces self-propulsion. 
What we have in mind are Janus particles~\cite{Walther2013,Wheat2010,Menzel2015,maggi2015} immersed in a solution   
(those spherical nano-particles are coated at their two hemispheres with a different materials).
Using a tight-binding language, the self-propulsion is incorporated as follows:   
a $w_n^{\nu}$ term is added to the $w_n^{+}$ rate (to the $w_n^{-}$ rate)   
if the particle faces the forward (backward) direction respectively. 
The particle flips orientation randomly with rate $w_n^{\alpha}$. 
\end{itemize}
We would like to treat both models on equal footing. In both models 
the intensity of the irradiation is denoted by~$\nu$. 
It is important to emphasize that the irradiation by itself does 
not create any bias (the rates are enhanced ``equally" in both directions).  
Given $\nu$ we can add an external bias~$f$, 
and ask whether a threshold value $f_c$ emerges.

\sect{Open issues}
The emergence of Sinai Physics due to {\em driving} or {\em propulsion} is somewhat counter-intuitive.
In quasi-equilibrium conditions, under the influence of a bath that has uniform temperature, 
the system feature normal diffusion, while the extra transitions due to driving or propulsion mechanism 
lead to sub-diffusion and localization.     
This (and similar) counter-intuitive statements have been pointed out in some previous works
in a way that was either blurred due to emphasis on related themes \cite{Hurowitz2013,Shapira2018}, 
or restricted to non-Brownian ``run-and-tumble particle" scenario \cite{dorkafri2019disorderactive,doussal2020runtumble}
where the motion is frozen in the absence of propulsion (rather than diffusive).   
Also the strong relation between tight-binding chains and their continuum limit version 
has not been adequately clarified.  In this context one should be aware that the standard 
theory of Sinai physics does not apply for non-equilibrium quasi-one-dimensional chains. 
This means that an active-particle, unlike passive-particle, is not a-priori compatible with Sinai model.   

\rmrkR{The experimental aspect has to be addressed carefully. 
One arena where 1D stochastic physics can be tested is in the context of Brownian Motors \cite{hanggi2009artificial}. 
The major signatures of Sinai physics is sub-diffusive spreading. 
In is common to determine the dependence of the sub-diffusion exponent on the bias, 
and to look for a sliding transition.            
A second arena for testing stochastic physics is inspired 
by experiments that we mentioned with Janus particles~\cite{Walther2013,Wheat2010,Menzel2015,maggi2015}. 
We claim that the realization of an appropriate setup 
for demonstrating Sinai Physics in this context is challenging (see concluding section). 
In such system it might be feasible not only to characterize the spreading, 
but also to probe the relaxation modes \cite{Shapira2018}, 
looking for a crossover from over-damped to under-damped relaxation (delocalization transition). }

\sect{Outline}
In \Sec{sec:rate} we review the basic notion that are used in order 
to describe stochastic dynamics that is generated by a rate equation.    
In \Sec{sec:sinai} we explain the essence of Sinai physics, 
and present some numerical results to motivate the later analysis.   
In particular we make the distinction between the non-equilibrium 
physics of active-particle and the quasi-equilibrium physics of passive-particle. 
In \Sec{sec:passive} and \Sec{sec:cont} we provide detailed analysis for passive-particle, 
both for its tight-binding version, and for its Fokker-Plank continuum limit. 
The dependence of $f_c$ on the driving intensity $\nu$ is found.  
%
%
In \Sec{sec:active-brownian} we consider the emergence of Sinai Physics for a Brownian active particle: 
without propulsion the particle performs diffusion, while with propulsion it becomes sub-diffusive.    
In \Sec{sec:deloc} we focus on the delocalization perspective. 
In particular we discuss the question whether for a finite size sample 
a residual affinity can induce spontaneous delocalization. 
Finally, in \Sec{sec:discussion} we summarize our findings, 
and critically discuss their relevance for experiments with Janus particles.

\section{Rate equations}
\label{sec:rate}

\rmrk{In this section we summarize the terminology that is used in order 
to mathematically-characterize a rate-equation $\dot{\bm{p}}=\bm{W} \bm{p}$  
that is defined over a network. 
In particular we define what are affinities, and what does it mean `detailed balance', 
as opposed to genuine non-equilibrium conditions. 
Subsequently we focus on (infinite) {\em Chain} and (finite) {\em Ring} geometries.}

\sect{\rmrk{Stochastic field}}
Consider a system that is described by a rate equation. From a mathematical point of view we can regard the set of sites 
as nodes of a network, labeled by some index $n$. On each bond of the network we have forward and backward transition rates, and the stochastic field is defined as 
\beq \label{Edef}
\mathcal{E}_{nm} \ \ \equiv \ \ \ln \left[\frac{w_{mn}}{w_{nm}}\right]
\eeq
If the rates are due to a bath that has a different temperatures $T_{nm}$ on each bond, we get 
\beq  \label{Enm}
\mathcal{E}_{nm} \ \ = \ \ \left[\frac{V_{n}-V_{m}}{T_{nm}}\right]  
\eeq
The circulations of $\mathcal{E}$ are called 
affinities. Using obvious continuum-limit notations
the affinity of a closed loop is   
\beq
\Phi \ \ \equiv \ \ \oint  \mathcal{E}(x) dx   
\eeq
For uniform temperature~$T$ any circulation of $\mathcal{E}$ is zero, 
meaning that the field is conservative. 
We than say that we have {\em detailed-balanced} conditions that reflect equilibrium.
\rmrk{More generally, even if the temperature is not uniform, 
we might have a {\em conservative} stochastic field. 
In particular, this is always the case for a one-dimensional chain 
with only near-neighbor transitions, simply because such geometry
has trivial topology with no circulations. }

\sect{\rmrk{Steady State}}
Whenever we have detailed-balanced \rmrk{(absence of non-zero circulations)} 
the stochastic field is conservative, 
and then we can define a stochastic potential~$U$ such that 
\beq \label{Un}
\mathcal{E}_{nm} \ \ = \ \ U_{n}-U_{m}   
\eeq
\rmrk{Consequently}, if the system is of finite-size,  
it relaxes towards a canonical steady state (SS):
\beq \label{eSS}
p_n^{\text{SS}} \ \ \propto \ \ e^{-U_n}  
\eeq
On the other hand, if the stochastic field is not conservative 
(absence of detailed balance due to non-zero circulations), 
we cannot derive it from a potential~$U$, and therefore the steady 
state is not canonical. Such non-equilibrium steady state (NESS) 
features non-zero currents.

\sect{\rmrk{Passive particle}}
Consider the passive particle of \Fig{f1}(a). For a {\em Chain} of length~$L$ one observes that due to the trivial topology (no circulations) the system is detailed-balanced.  This holds even if each of the bonds has a different temperature. Adding a driving source, the potential~$U$ is still well defined. What we call bias refers to the average value of the stochastic field:
\beq \label{fdef}
f \ \ \equiv \ \ \frac{1}{L} \sum_n \mathcal{E}_n 
\eeq
For a {\em ring} of length $L$, non-zero bias implies non zero affinity ${\Phi = f L}$. 
For zero affinity (${\Phi=0}$) the potential~$U$ along the ring is well defined, 
and the NESS is canonical. For a biased ring the NESS features a non-zero current.
An explicit expression that generalizes \Eq{eSS} can be found in \App{sec:drift-derrida}.

\sect{\rmrk{Active particle}}
Considering the active particle of \Fig{f1}(b), one observes that there is non-zero circulation in each cell of the network. The nodes of the network are labeled by the coordinates $(n,s)$, where $n$ is the site index, and ${s=\uparrow,\downarrow}$ indicates the forward/backward orientation of the particle. The circulation around a unit cell reflects the propulsion. We are dealing here with a genuine non-equilibrium system. It is not possible to define a stochastic potential~$U$. 
An exact explicit expression that generalizes \Eq{eSS} cannot be found, 
though approximations can be attempted in limiting cases, e.g. see \App{sec:active-non}.

\section{The emergence of Sinai Physics}
\label{sec:sinai}

In this section we clarify how `Sinai Physics' emerges in the context
of an (infinite) Chain and in the context of a (finite) Ring.  
In particular we illuminate subtleties that are related to the identification 
of Sinai Physics for active particles.

\sect{Sinai physics for a passive particle}
Sinai Physics arises for the passive particle of \Fig{f1}(a) 
if the $\mathcal{E}_n$ become effectively uncorrelated.  
What we mean by `effectively uncorrelated' is that the 
stochastic potential $U$, unlike $V$, becomes unbounded  
such that ${ |U(x+r)-U(x)|  \sim \sqrt{r} }$.
%
The simplest way to get an effectively uncorrelated potential 
is to consider hypothetical situation where each bond has a different temperature. Such type of disorder breaks the telescopic correlations of stochastic field \Eq{Enm}.  The integral over this field provided the stochastic potential of \Eq{Un}. This integral is formally like adding random variables, and therefore features $\sqrt{L}$ fluctuations over segments of length~$L$. Those unbounded fluctuations are responsible for the sub-diffusive spreading that has been highlighted by Sinai and followers.        

\sect{Sinai physics for an active particle}
The standard argument that has established the emergence of Sinai physics for a passive particle, fails for the active particle of \Fig{f1}b. Namely, in the latter case it is impossible to define a stochastic potential~$U$. Still we can argue that this quasi-one-dimensional system is similar to Sinai model, namely, it can be approximated by the same geometry as that  of \Fig{f1}a. We shall discuss 3 options: 
{\bf (1)}~Reduction to an effective Sinai model; 
{\bf (2)}~Inspection of the NESS;
{\bf (3)}~Inspection of the spectrum. 
\rmrk{The first option would provide, at best, an approximation. 
We shall see that in practice it is quite limited, and not very satisfactory.   
Let us elaborate further on the two others options.}

\sect{Inspection of the NESS}
\rmrk{One can argue that the failure to find a satisfactory reduction of a quasi-one-dimensional network to an effective single-channel Sinai-model is a technical issue. Maybe in principle it is always possible to find an effective-Sinai-model? Let us see what is the implication of such {\em conjecture}, and whether we can benefit from it.} Assuming that the active particle indeed can be approximated, after coarse-graining, by an effective Sinai model, it follows that the NESS is canonical-like. Then it is possible to extract an effective stochastic potential via the NESS. \rmrk{Namely, assuming that the NESS is provided by \Eq{eSS} with an effective potential $U_n$, it follows that the effective field is:}  
\beq \label{e7}
\mathcal{E}_{n}^{\text{eff}} \ \ \equiv \ \ \ln \left[ \frac{ p_{n{+}1}^{\text{NESS}} }{ p_n^{\text{NESS}} } \right]
\eeq
where ${p_n = p_{n,\uparrow} +  p_{n,\downarrow} }$ is the probability to find the particle in site~$n$.
\rmrk{For the standard Sinai model the above definition leads to the trivial identification 
${ \mathcal{E}_{n}^{\text{eff}} = \mathcal{E}_{n} = \ln(w_n^{+}/w_n^{-}) }$. 
But for a quasi-one dimensional network, that does not satisfy the detailed-balance condition, 
there is no simple way to relate the effective field to the local rates. 
Still, from a practical point of view we can benefit from the above definition. 
Given $\bm{W}$, it is a rather easy task to find the NESS, and using \Eq{e7} 
to deduce ${ \mathcal{E}_{n}^{\text{eff}} }$.   
The effective stochastic potential $U_n^{\text{eff}}$ is obtained by integrating over the field. 
From this effective potential we can deduce what is~$f_c$. 
We will test {\em numerically} whether this {\em conjecture-based} prediction is valid.}   
%

\begin{figure}
\centering
\includegraphics[]{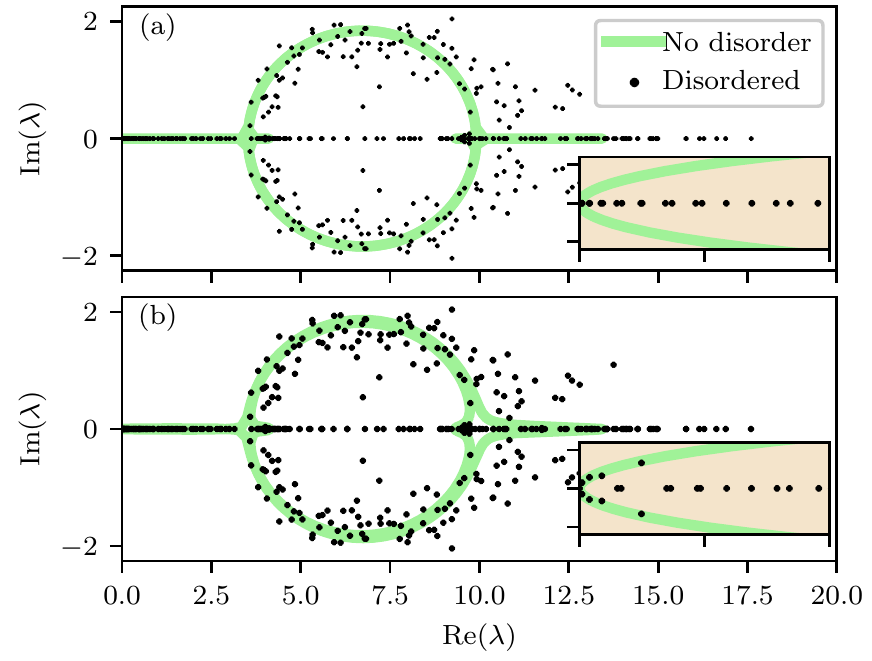} 
\caption{
The relaxation spectrum for the active particle of \Fig{f1}b.
The thick lines are $\lambda_{k,\pm}$ of the two bands via 
exact diagonalization of $\bm{W}$ for a non-disordered system.
The dots are found via numerical diagonalization for a ring of length ${L=300}$. 
The unbiased transition rates are $w^{\beta}{=}1$,
the propulsion rate is $\ln(w^{\nu}){=}1$, 
and the flipping rate is $w^{\alpha}{=}2$. 
The disorder parameters are ${\sigma_{v}=1}$ and ${\sigma=0.1}$, 
as defined after \Eq{eq:fc}.
Panels~(a) and~(b) are for ${f=10^{-4}}$ and ${f=10^{-2}}$,
respectively, that are smaller and larger than~$f_c$. 
The insets depict the spectrum at the vicinity of ${\lambda=0}$, 
with a zoom of factor $\times 10^4$ in~(a) and $\times 10^2$ in~(b) for the vertical axis, 
and $\times 3$ for the horizontal-axis.
For a non disordered system the zoomed region 
can be approximated by \Eq{eSpectrum} where ${v \propto f}$.
With disorder the spectrum in the vicinity of ${\lambda=0}$ remains real for ${f<f_c}$,   
and indicates delocalization for ${f>f_c}$.   
\label{f2}}
\end{figure}

\sect{Inspection of the Spectrum} 
\rmrk{The question arises how to test predictions for $f_c$ that are based on a coarse-graining {\em approximation} or on a {\em conjecture} regarding its feasibility. We ask for an approximation-free  conjecture-independent way to determine~$f_c$. To determine $f_c$ numerically for the sliding transition is not practical (as explained below). Fortunately, Sinai physics can be detected via the delocalization transition that precedes the sliding transition.} 
This option is extremely convenient numerically for the following reason: Numerical testing of the sliding transition is too tough, because it requires a large ensemble and proper finite-size-scaling procedure. The drift velocity~$v$ is zero for ${f<f_c}$ only if the ${L\rightarrow \infty}$ limit is taken, and an average over realizations is formally required. In contrast, the related delocalization transition is easy for detection. Considering a single realization of a ring that has a finite length~$L$, one can look for the threshold field that is required to get complex eigenvalues at the vicinity of ${\lambda=0}$. For a clean ring, that has the simple geometry of \Fig{f1}a, the  eigenvalues become complex for any ${f>0}$, namely, 
\beq \label{eSpectrum}
\lambda_k \approx D k^2 + i vk, 
\ \ \ \ \  \
k=\frac{2\pi}{L} \times \text{integer}   
\eeq
where $v$ is the drift velocity, and~$D$ is the  diffusion coefficient. With disorder this approximation is valid only if both $v$ and $D$ are finite. This happens for a bias that {\em exceeds} the critical value that is calculated for the sliding transition. But a proper analysis \cite{HurowitzCohen2016} has clarified that the actual threshold for the delocalization transition {\em precedes} the estimated value for the sliding transition. We shall further discuss those thresholds in \Sec{sec:deloc}.

For the active particle of \Fig{f1}b the delocalization transition is more subtle. Here complexity of an eigenmode  does not imply that it is {\em extended} in space. The matrix $\bm{W}$ is not similar to an hermitian matrix even if the bias~$f$ is zero, since the hermiticity is spoiled by the propulsion. Complex eigenmodes can reside in localized loops of the network. 
Still, we can rationalize the association between the sliding transition and delocalization as follows. 
In the absence of disorder, and for zero propulsion, 
the spectrum is real and consists of two bands, see \cite{Shapira2018} and \App{sec:diffusion-active}. 
The lower band extends from ${\lambda=0}$ up to twice the switching rate, 
where it start to overlap with the second band. 
Complexity due to propulsion appears in the band-overlap region, 
and does not affect the ${\lambda=0}$ region.   
(a detailed explanation of this point is provided at the end of \App{sec:diffusion-active}).
It is therefore valid to claim that features of the spectrum 
at the vicinity of ${\lambda=0}$ provide indication for the coarse 
grained dynamics of the spreading. We therefore expect a Sinai-type 
scenario. This expectation is supported by the demonstration in \Fig{f2}.

\section{Passive particle}
\label{sec:passive}

In this section we consider the emergence of Sinai physics for the
chain whose unit cell is sketched in \Fig{f1}a. 
We treat an $L$ site ring and a chain on equal footing. 
Accordingly we assume periodic boundary conditions, 
and label the sites by $n$ modulo $L$, or by $x$ in the continuum limit. 
The forward(+) and backward(-) transition rates between the $n$ and $n{+}1$ sites are:
\beq \label{eq:transition-rates}
w_{n^{\pm}} \ \ &=& \ \  w_n^{\beta}  e^{\pm \Delta_n/2 T_n} \ + \ w_n^{\nu} \
\\ \label{eq:transition-rates-a}
\ \ &\equiv& \ \ w_n^{\beta} \left[  e^{\pm \Delta_n/2 T_n} + \nu g_n \right]
\\ \label{eq:transition-rates-b}
\ \ &\equiv& \ \ w_n^{\beta} \ e^{\pm \mathcal{E}_n/2} 
\eeq
The first term in \Eq{eq:transition-rates}
represents bath induced transitions between sites 
that have a potential difference ${ \Delta_n=-[V_{n{+}1}-V_n] }$.
The local bath temperature is $T_n$. The second term  $w_n^{\nu}$ 
represent extra transitions that are controlled 
by an illumination source of intensity~$\nu$. 
The optional notation of \Eq{eq:transition-rates-a} defines 
dimensionless coupling $g_n$. In the numerics we assume that thy have a log-normal distribution.  It is convenient to normalize them such that ${\braket{\ln(g_n)} =0 }$, and thus the definition of the irradiation intensity $\nu$ is implied.  
Finally, the stochastic field on the $n$-th bond has been  
defined in accordance with \Eq{Edef}, namely,  
\beq \label{EdefChain}
\mathcal{E}_n \ \ = \ \ \ln \left[\frac{w_{n^{+}}}{w_{n^{-}}}\right]
\eeq
The high-temperature approximation can be written as 
\beq \label{eBeta}
\mathcal{E}_n \ \ &\approx& \ \ \left( \frac{1}{1+\nu g_n} \right) \left[\frac{V_{n{+}1}-V_n}{T_n}\right]
\\ \label{eBeta1}
\ \ &\equiv& \ \ -\beta_n  [V_{n{+}1}-V_{n}] 
\eeq
\rmrk{where the $\beta_n$ are defined via the last equality.}
The average stochastic field~$f$ has been defined in \Eq{fdef}.
If the only mechanism for transition is a bath with uniform temperature we get ${f=0}$. 
We can get  ${f \ne 0}$ either due to non-uniform $\beta_n$, or by adding an external bias field,
aka affinity (the former term is commonly used for a chain, while the latter is commonly used for a finite ring).  
The practical way to introduce bias~$f$ 
in a numerical illustration is to use the prescription
\beq 
w_{n^{\pm}} \ \ \mapsto \ \ w_{n^{\pm}} \exp({\pm f/2}) 
\eeq
In accordance with this reasoning 
we define an $L$ periodic stochastic potential $U_n$ for zero bias, 
while for finite bias we write    
\beq
\mathcal{E}_n  \ \ \equiv \ \ f - [U_{n{+}1}-U_{n}]
\eeq

\sect{The drift velocity}
Due to $f$ one obtains NESS with non-zero current that has a drift velocity  \cite{Derrida1983,BOUCHAUD1990,BOUCHAUD1990a} 
\beq \label{vdrift}
v_{\text{drift}} = (1{-}e^{-f L})
\left[ \frac{1}{L}\sum_{n=1}^{L} \frac{1}{w_{n^{+}}} \sum_{r=0}^{L{-}1} e^{- fr + [U_{n} {-} U_{n-r}}] \right]^{-1} 
\eeq
The infinite chain limit is formally obtained by setting ${L=\infty}$.
In the absence of disorder one obtains an $L$-independent drift ${v_{\text{drift}} = w_{+} \left( 1 - e^{-f} \right) = w_{+}-w_{-}}$, and for weak bias
${ v_{\text{drift}} = w f }$, as expected.
But if the stochastic field is disordered the drift velocity vanishes for ${f<f_c}$.  The definition of the critical field $f_c$ is implied by inspection of the sum in the square brackets: 
by definition the sum diverges for ${f<f_c}$.

\sect{Uncorrelated field}
Let us first remind how $f_c$ is estimated for the standard Sinai-Derrida model, where the $\mathcal{E}_n$ are assumed to be uncorrelated random variables. In the absence of bias ($f{=}0$) we define ${ \braket{e^{-\mathcal{E}}} \equiv e^{f_1} }$. 
Then, in the presence of finite bias $f$ we get 
\beq \label{eq:v-drift}
v_{\text{drift}} \ \ &\approx& \ \ 
\left[ \frac{1}{L}\sum_{n=1}^{L} \frac{1}{w_{n^{+}}} \sum_{r=0}^{L-1} e^{- (f-f_1)r} \right]^{-1} 
\\
\ \ &=& \ \  \braket{ \frac{1}{w_{n^{+}}} }^{-1} \, \left[1 - e^{-(f-f_1)}\right]
\eeq
where the latter equality holds if the geometric sum converges, i.e. for ${f>f_1}$. Accordingly one deduces that ${f_c=f_1}$. 
If the stochastic field has a Gaussian distribution we get ${f_c=f_1=(1/2)\text{Var}(\mathcal{E})}$.

\begin{figure}
\centering
\includegraphics[]{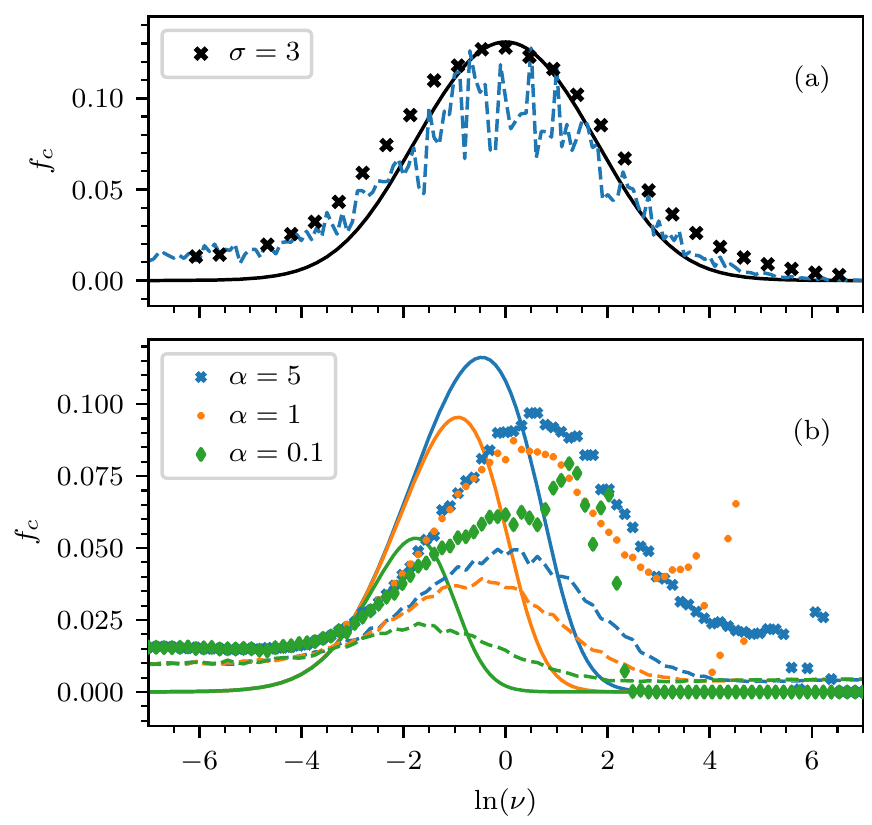} 
\caption{
The threshold field $f_c$ versus the irradiation intensity~$\nu$.
The numerics is done for a ring of length ${L=200}$ with $w^{\beta}{=}1$.
The disordered potential is with ${\sigma_{v} = 2}$ and $\sigma{=}3$. 
\textbf{(a)}~A passive particle:
The analytical estimate for $f_c$ (solid lines) is based on \Eq{eq:fc}. 
It is compared with a numerical estimate that is based on $50$ realizations 
using \Eq{eq:fc-varu} with \Eq{EdefChain} (dashed lines).    
For the same realizations we numerically determine 
the average delocalization threshold (symbols).
%
\textbf{(b)}~An active particle:
The analytical estimates for $f_c$ (solid lines) 
are obtained from \Eq{eq:fc} using $\beta_n$ of \Eq{eq:beta-x-active} 
(see text for further details and discussion). 
The NESS based estimates (dashed lines) are based (each) on $2000$~realizations
using \Eq{eq:fc-varu} with the field of \Eq{e7}.   
Both estimates are compared with the numerically determined average delocalization threshold (symbols).
The solid and dashed lines in both panels are divided by 3 as explained in \Sec{sec:deloc}.
\label{f3}}
\end{figure}

\sect{Correlated field}
In our case the $\mathcal{E}_n$ are correlated, and not Gaussian distributed.  We looks on the random variable ${u = (U_{n{+}r} - U_n) }$. 
It has zero average, and we assume that it is ``small". 
The smallness holds because our interest is in the high temperature regime, which is effectively like a weak disorder assumption. Furthermore we focus on segments with range~$r$ that are larger by a  comparable factor with the spatial correlation scale. For such segments one can use the estimate  ${\braket{\exp(u)} \approx \exp[(1/2)\braket{u^2}]}$. Longer segments can be ``factorized" as in the treatment of uncorrelated field. 
The bottom line is that we can extract the critical field through the approximation 
\beq \label{eq:fc-varu}
\dfrac{1}{2}\braket{u^2} \ \ \approx \ \ f_c \, r
\eeq
where $f_c$ is defined as the prefactor in this linear dependence.
Specifically we assume that the stochastic field is given by \Eq{eBeta1}. 
The potential $V_n$ is assumed to be uncorrelated, with zero average.
\rmrk{The random variable ${u = (U_{n{+}r} - U_n)}$ is the sum ${ \sum \mathcal{E}_{n'} }$ 
of correlated variables ${ \mathcal{E}_{n'} = \beta_{n'} [V_{n'{+}1}-V_{n'}]}$.
This can be re-arranged as a sum over uncorrelated variables. Namely,
neglecting a boundary term, ${u := \sum [\beta_{n'}-\beta_{n'{-}1}] V_{n'}}$.  
Therefore we deduce that $\braket{u^2}$ equals the number $r$ of terms in the sum, 
times the variance ${\braket{ [\beta_{n}-\beta_{n{-}1}]^2 V_{n}^2} }$ of each term. 
Using ${\braket{ [\beta_{n}-\beta_{n{+}1}]^2} = 2\mbox{Var}(\beta_n)}$, 
it follows from \Eq{eq:fc-varu} that the critical bias is} 
\beq \label{eq:fc}
f_c \ \ \approx \ \ \text{Var}({\beta_n}) \, \text{Var}(V_n^2) 
\eeq  
By inspection of \Eq{eBeta}, non zero $\text{Var}({\beta_n})$ is implied either by random local temperature $T_n$, or by the presence of a driving source $\nu$.
The latter mechanism assumes that there is some bond-disorder (the bonds are not identical), or optionally a non-uniform irradiation.
To obtain a concrete result, we choose $(V_n/T)$ to be normally distributed with variance $\sigma_{v}$, and $g_n$ to have a log-box distribution, $\ln(g_n) \sim [-\sigma,\sigma]$. The critical field is then
\beq
f_c \ = \ 
\dfrac{\sigma^2_{v}}{4} \left( 
1   -  \dfrac{1}{\sigma^2}\log^2 \left[  \dfrac{\cosh ((s+\sigma)/2)}{\cosh ((s-\sigma)/2)} \right]  
-\dfrac{2}{\sigma}\dfrac{ \sinh(\sigma)}{\cosh(\sigma) + \cosh(s)}
\right) 
\label{eq:fc-explicit}
\eeq
with $s \equiv \ln(\nu)$. This expression has a bell like shape, 
with a maximum at $s=0$, for which
\beq \label{fcmax}
f_c \ = \ \left( 1  - \frac{2}{\sigma} \tanh \left( \dfrac{\sigma}{2} \right) \right)  
\frac{\sigma_{v}^2}{4} 
\eeq
We numerically illustrate this result in \Fig{f3}a (solid line). 
In this figure also the threshold field for delocalization is displayed.
The latter is expected to be the same up to a factor 
(we further discuss this issue in \Sec{sec:deloc}).   
Without any external field, the spectrum is real.
The threshold field for delocalization is determined
by the appearance of complex eigenvalues near ${\lambda=0}$.
The result for $f_c$ is averaged over many realizations.
%
%
We shall see later that the propulsion mechanism has effectively 
a similar effect as illustrated in \Fig{f3}b.

\section{Passive particle - continuum version}
\label{sec:cont}

We consider the continuum version of the model that is sketched in \Fig{f1}a.
In this limit the derivation of $f_c$ becomes more illuminating.  
The continuum version of ``random walk" on a lattice is 
a Fokker-Planck equation that includes diffusion and drift terms: 
\begin{align}\label{eq:fp}
\dfrac{\partial \rho}{\partial{t}} &=
-\dfrac{\partial}{\partial{x}} \left( v(x)  \rho -D(x)\dfrac{\partial \rho}{\partial x} \right)
\end{align}
This equation can be regarded either as a stand-alone model, 
or as an approximation for the lattice model. 
In the absence of disorder the diffusion coefficient 
is \rmrk{${D_0 \equiv (1/2) (w^{+}+w^{-})a^2 }$}, 
and the drift velocity is \rmrk{${v = (w^{+}-w^{-})a }$}, 
where $a$ is the lattice constant (set as unity in the previous section).  
In the presence of disorder the drift and diffusion terms are:
\beq \label{eD}
D(x) &=&  \left( 1 + \nu g(x) \right) D_0 
\\ \label{ev}
v(x) &=& - \frac{1}{T} V'(x) D_0    \ \ \equiv \ \ - \mu_0 V'(x)
\eeq
where $V(x)$ and $g(x)$ are the continuum limit version of $V_n$ and $g_n$. 
In complete analogy with the tight-binding version we define 
\beq \label{e22}
\beta(x)  &\equiv&  \left(\frac{1}{1 + \nu g(x)}\right) \frac{1}{T} \\
\mathcal{E}(x) &\equiv&   \frac{v(x)}{D(x)}  \ = \ -\beta(x) V'(x) \ \ \equiv \ \ -U'(x)
\eeq
The role of the lattice constant~$a$ is played by a correlation distance.
We define 
\beq
C_V(r) &=& \avg{V(x)V(x+r)} \\
C_{\beta}(r) &=& \avg{\beta(x)\beta(x+r)} - \avg{\beta(x)}^2 
\eeq
For Gaussian distribution we write 
\beq
C_V(r) &=& \text{Var}(V) \, e^{- \frac{1}{2} \left( \frac{r}{a} \right)^2} \\
C_{\beta}(r) &=& \text{Var} (\beta) \, e^{-\frac{1}{2} \left( \frac{r}{a} \right)^2}
\eeq

The continuum version of the Derrida formula \Eq{vdrift}, 
whose derivation is detailed in \App{sec:drift-derrida}, 
takes the form 
\beq \label{vdriftc}
v_{\text{drift}} = (1 {-} e^{-f L})
\left[ \int_0^L dx \braket{ \frac{1}{D(x)} e^{-f x + U(x) }} \right]^{-1} 
\eeq
In this version an ensemble average has been incorporated, 
and the convention ${U(0)=0}$ is implicit. 

The derivation of an expression for $f_c$ proceeds 
as in the previous section. Using 
\beq \nonumber
&& \braket{U(x)^2} =  
\int_{0}^{x} \!\!\! \int_0^x   dx_1 dx_2 \avg{ \beta(x_1)\beta(x_2) V'(x_1)'V(x_2)}  
\\ \nonumber
&& \ \ \ = x  \int_{-\infty }^{\infty} dr \left(C_{\beta}(r) + \avg{\beta(x')}^2  \right) \left( - \dfrac{d}{dr^{2}} C_V(r) \right) \\
&& \ \ \ = x \int_{-\infty }^{\infty } dr\, C'_\beta(r) C'_V(r) 
\eeq
we get 
\beq \label{eq:fcc}
f_c \ \ = \ \ \frac{\sqrt{\pi}}{4a} \text{Var}(\beta) \, \text{Var}(V)
\eeq
This result is consistent with \Eq{eq:fc},
where we assumed Kronecker delta correlations,  
and lattice-constant that equals unity.

\section{Active Brownian particle}    
\label{sec:active-brownian}

We turn now to discuss the active particle system of \Fig{f1}b. 
Unlike the non-Brownian version that has been discussed in the past (see \App{sec:active-non}),  
here we have diffusion to begin with, and we ask what 
happens if we add propulsion. The rates of transitions are:    
\beq
w_{(n{+}1) \uparrow, n \uparrow} &=&   w_n^{\beta} e^{+\Delta_n/2T_n} + w^{\nu}_n \\
w_{n \uparrow, (n{+}1) \uparrow} &=&   w_n^{\beta} e^{-\Delta_n/2T_n} \\
w_{(n{+}1) \downarrow, n \downarrow} &=&  w_n^{\beta} e^{+\Delta_n/2T_n}  \\
w_{n \downarrow, (n{+}1) \downarrow} &=&   w_n^{\beta} e^{-\Delta_n/2T_n} + w^{\nu}_n \\
w_{n \downarrow, n \uparrow} \ \ \ \ \ &=& w_n^{\alpha} \\
w_{n \uparrow, n \downarrow} \ \ \ \ \ &=& w_n^{\alpha}
\eeq
The diffusion coefficient in the absence of disorder is
calculated in \App{sec:diffusion-active}. The final result is
%
%
%
%
\beq \label{eDeff}
D^{\text{eff}} \ \ = \ \  w^{\beta}a^2 + \frac{1}{2}w^{\nu}a^2 + \frac{[w^{\nu}]^2}{2w^{\alpha}}a^2  
\eeq
All 3 terms in $D_{\text{eff}}$  have a simple heuristic explanation. 
The last term is consistent with the continuum limit \Eq{e31}.
Note that the continuum limit (${a \rightarrow 0}$) is taken 
such that ${w^{\beta}a^2=D}$ and ${w^{\nu}a=\nu}$ 
are kept constant. Therefore the second term drops out. 
The first term is excluded if one considers the non-Brownian version.   
Once we have disorder it is convenient to define local drift velocity 
and local diffusion coefficient as follows:
\beq
v_n &=& \left[2\sinh\left( \frac{\Delta_n}{2T_n} \right)\right] \, w_n^{\beta}a \\
D_n &=& \left[\cosh\left( \frac{\Delta_n}{2T_n} \right)\right] \, w_n^{\beta}a^2 
\eeq
The zero order effect of the disorder 
is to replace the first term in \Eq{eDeff} by an averaged $D_n$. 
Such zero-order treatment has no profound meaning 
and merely reflects our definition of $w^{\beta}$ 
as geometric average and not as the algebraic average 
of the backward and forward rates.

\sect{NESS based approach}
In \App{sec:ness-active} we find the NESS, based on a non-Brownian approximation.  
Namely, we neglect counter-propulsion transitions, and assume weak disorder.  
Using units such that ${a=1}$ the result is 
\beq
\frac{p_{n{+}1}}{p_{n}} 
\ \ \approx \ \ 
\left[ \frac{\alpha + \nu - v_n}{\alpha + \nu + v_n} \right]   
\left(\frac{\nu+v_n}{\nu-v_n} \right)
\eeq
Weak disorder means that the ${v_n }$ are much smaller than the propulsion velocity~$\nu$. 
An expression for the effective $\mathcal{E}_n$ is implied via \Eq{e7}.
The first order expression in $v_n$ is 
\beq
\mathcal{E}_n \ \ \approx \ \ \frac{2\alpha v_n}{(\alpha+\nu)\nu} 
\eeq
Going beyond linear order but assuming small $\alpha$ 
we get a result that is consistent with the continuum limit
\beq
\mathcal{E}_n \ \ \approx \ \ \frac{2 \alpha v_n}{\nu^2-v_n^2}
\eeq
Either way we can write ${ \mathcal{E}_n \equiv \beta_n v_n }$, 
where $\beta_n$ might reflect disorder that is related to $\alpha$ or $\nu$.
In the absence of such disorder the higher order dependence
on $v_n$ has to be taken into account.       
Having disordered $\beta_n$ is required for the emergence of Sinai physics, 
as explained in previous sections.

As far as analytical estimates are concerned, the NESS based approach to obtain the stochastic field, via \Eq{e7}, is a dead end. Specifically, for an active Brownian particle, there is no simple way to generalized the solution such that it will take into account the counter-propulsion rates. 
We therefore adopt in the remaining part of this section a more flexible heuristic approach for the estimate of $f_c$.
Numerically we compare the two methods in \Fig{f3}. 
Namely, we estimate $f_c$ from the numerically found NESS, 
and compare it with the heuristic analytical estimate.  
In the next section we highlight a somewhat more rigorous approach.

\sect{Heuristic approach}
The heuristic approach is based on the assumption that 
the {\em coarse grained} dynamics is described by a single-channel diffusion equation, 
namely \Eq{eq:fp}. 
It is therefore natural to adopt continuum-limit notations.
Within this framework Sinai physics emerges whenever we break locally the Einstein relation, namely,  
\beq
\frac{\mu(x)}{D(x)} \ \ = \ \ \beta(x) \ \ = \ \ \text{disordered}
\eeq
Then the critical field is determined by \Eq{eq:fcc}.
For ${\nu=0}$ the local mobility $\mu_0$ 
and the local diffusion coefficient $D_0$ 
are defined as in \Eq{ev} and \Eq{eD}.
More generally both may depend on~$x$,
but this dependence is not important for us 
because the ratio ${\mu_0/D_0=1/T}$ 
does not care about this dependence. 
For ${\nu \ne 0}$ the mobility is barely affected 
because the extra transitions do not bias the transitions.
This is true for both passive and active particles. 
So we can write ${\mu(x) = \mu_0(x)}$. 
But $D(x)$ is affected. 
For a passive particle see \Eq{eD}.
For an active particle we can justify 
the use of \Eq{eDeff} (with an added site coordinate) 
provided the variation of $V(x)$ is on large scale ${a \gg \text{LatticeConstant}}$.
But in practice we test in \Fig{f3} our formulas 
for uncorrelated potential (${a =  \text{LatticeConstant} }$), 
so inaccuracies are expected.

From the above discussion it follows that for the purpose 
of estimating $f_c$ via \Eq{eq:fcc} we use expression of the form
\beq \label{eq:beta-x-active-D}
\beta(x)  =  \frac{ \mu_0(x) }{D_0(x) + \nu D^{(1)}(x) + \nu^2 D^{(2)}(x) } 
\eeq
This expression with ${D_0=0}$ holds for non-Brownian particle, 
while for Brownian particle it is more illuminating to write it 
in the style of \Eq{e22}, namely, 
\beq \label{eq:beta-x-active}
\beta(x)  =  \frac{\beta_0(x)}{1 + \nu g^{(1)}(x) + \nu^2 g^{(2)}(x) }
\eeq
where $\beta_0(x)=1/T$ for a bath that has uniform temperature.

\sect{Smoothing effect}
In the numerics we set the units such that ${T=a=1}$, 
and use \Eq{eq:beta-x-active} with the substitutions  
\beq
g^{(1)}(x) \ &\mapsto& \ g_n  \\
g^{(2)}(x) \ &\mapsto& \  \text{Smoothed} \left\{ [g_n]^2/(2\alpha) \right\} 
\eeq
Then, $f_c$ is calculated using \Eq{eq:fc}.
The results are presented in \Fig{f3}b.
If we do not perform spatial smoothing there is a rough agreement 
with the delocalization numerics, provided $\alpha$ is large, 
and $\nu$ is not too large. 
Apparently we cannot expect better because the error  
is comparable with the difference between the delocalization-based 
result and the NESS-based result.  
But for small values of $\alpha$ one observes (not displayed) a gross over-estimate:  
the estimated $f_c$ becomes larger instead of getting smaller.
This over-estimate can be avoided if spatial smoothing is performed. 
The reasoning is as follows: if $\alpha$ is small the mean free path 
is large (${\ell \sim 1/\alpha}$). Therefore the effective 
diffusion term has to be smoothed over this scale. 
But alas, the smoothing procedure is analytically ill defined, 
and therefore in practice we were satisfied 
by the replacement ${ g^{(2)}(x) \mapsto 1/(2\alpha) }$. 
This substitution provides a lower bound for our estimate, 
hence becoming an under-estimate for large~$\nu$. 
However, for large $\nu$ there is a bigger issue that we discuss next.

\sect{Finite size effect}
We observed in \Fig{f3}b that the $f_c$ analytical estimate for the delocalization threshold fails for large $\nu$. The actual threshold drops to zero at a $\nu$ value that depends on $\alpha$. The failure to predict this drop cannot be attributed to its inherent limitations (those that have been pointed out in the previous paragraph).     
Rather some further inspection reveals that this drop of $f_c$ is a finite-size effect. The $\lambda_{k}$ spectrum for a non-disordered ring can be calculated analytically, see \App{sec:diffusion-active}. For a large enough ring the spectrum for $f{=}0$ is always real in the vicinity of $\lambda_0{=}0$.  More precisely the range where it is real 
is restricted by the condition  ${\sin(k) < \alpha/\nu}$. No eigenvalues reside within this range if $L$ is too small. Accordingly, Sinai physics is observed only for ${ \nu < \alpha L}$.

\section{The delocalization transition}
\label{sec:deloc}

So far we have regarded the delocalization transition 
as a numerical trick to detect the emergence of Sinai Physics.
But also from a physics-oriented perspective we have to remember 
that we are always dealing with a \textit{finite} system,
where it is meaningless to look for a sliding transition.
Namely, for a finite-length ring,  
a non-zero~$f$ always implies a non-zero drift velocity.
Thus, in the context of a finite-length ring 
it is appropriate to consider the delocalization transition
rather than a sliding transition. 
We start with the benchmark Sinai model, and then apply the same 
reasoning for the irradiated passive particle. 
The same reasoning can be applies also for an active particle.

\begin{figure}
\centering
\includegraphics[]{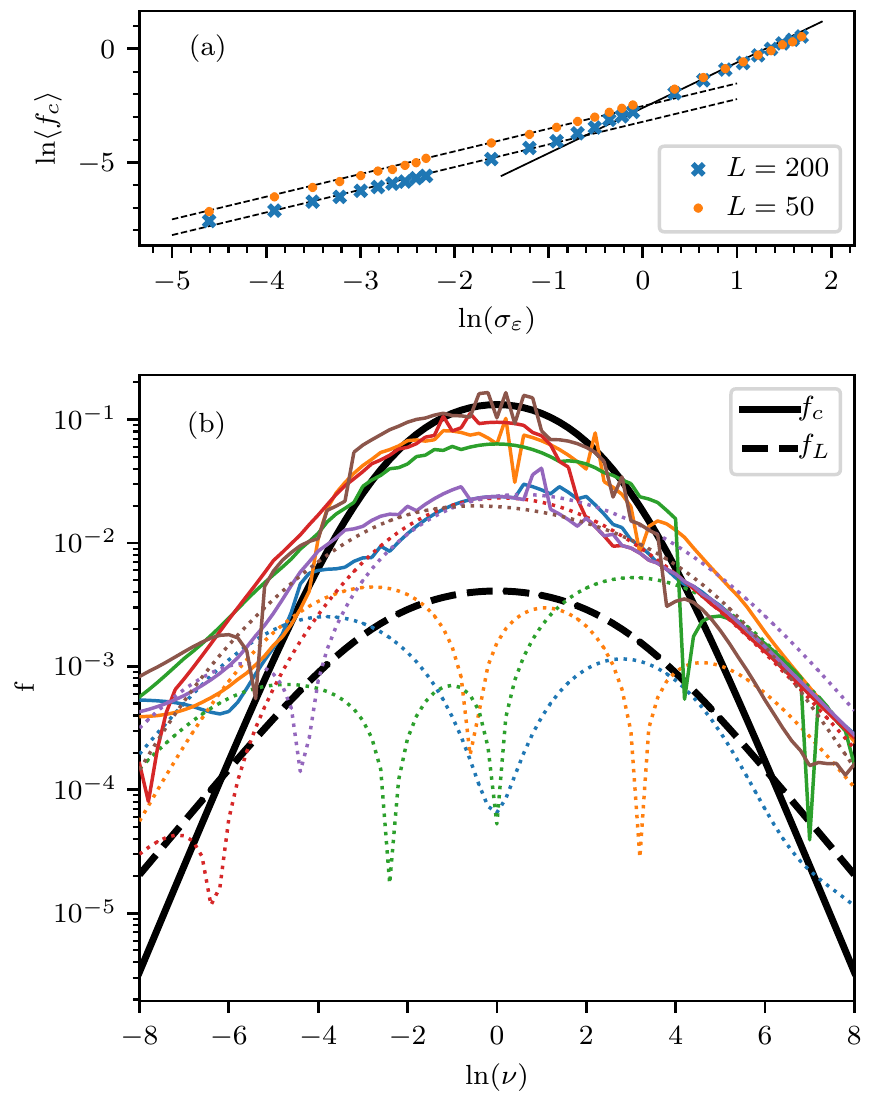}
\caption{ 
Demonstration of spontaneous delocalization.
\textbf{(a)~Benchmark Sinai model:} 
The average $f_c$ for uncorrelated fields, for two rings with different lengths.
Each symbol is the average of $f_c$ over $300$ realizations.
For ${f_c > f_L}$, the delocalization threshold
is ${f_c \approx f_{1/2} \propto \sigma_{\varepsilon}^2}$ (solid line),
while for weaker bias ${f_c \approx f_L \propto \sigma_{\varepsilon}  }$ (dashed lines), 
see \Eq{eq:fc-small-sigma}.
%
%
\textbf{(b)~Passive irradiated particle:}~
We consider an ${L \approx 2000}$ ring 
with the same disorder parameters as that of \Fig{f3}a. 
We plot the numerically-determined residual bias $f_L$ (dotted lines), 
and delocalization threshold $f_c$ (solid lines),
for a few realizations (distinguished by color).
The black thick solid and dashed lines   
have the same meaning as in panel~(a), 
namely, they correspond to $f_{1/2}$ and $f_L$.
Specifically, the black thick solid line is the same as in \Fig{f3}a. 
\label{f4}
}
\end{figure}

\sect{Benchmark Sinai model}
Consider a simple ring of length~$L$, 
with independent stochastic fields $\mathcal{E}_n + f$, 
and $f$ is defined such that ${\sum_n \mathcal{E}_n =0 }$, 
while $\var{(\mathcal{E})} = \sigma_{\varepsilon}^2$. 
We can also define the $\mu$-moment $f_{\mu}$ 
of the random stochastic field via 
\beq \label{eq:f-mu}
\braket{ e^{-\mu \mathcal{E}} } \equiv  e^{ -\mu f_{\mu}}
\eeq
For Gaussian disorder ${ f_{\mu} = (\mu/2) \sigma_{\varepsilon}^2  }$.
The bias $f$ is regarded as a control parameter, 
and the relation ${f=f_{\mu}}$ can be inverted.
Thus we can regard $\mu(f)$ as an optional way 
for characterization of the bias. 
Recall that the bias $f_c=f_1$, 
that corresponds to $\mu{=}1$,  
is the critical field for the sliding transition.

\sect{Delocalization threshold}
The sliding threshold~$f_1$ concerns an ensemble average over infinite chains,
while the delocalization threshold concerns a single realization of a finite size ring.
The relation between the two transitions has been studies in \cite{HurowitzCohen2016} 
for the benchmark Sinai model. 
Assuming a very long ring, the threshold field for sliding is determined 
by a spectral density $\lambda^{\mu-1}$, where ${\mu=\mu(f)}$.   
It has been argued that the delocalization transition takes place 
for ${\mu=1/2}$, meaning that ${f_c[\text{delocalization}] \approx f_{1/2}}$.
For Gaussian disorder ${ f_{1/2} =(1/2) f_1 }$, while for (e.g.) a box distribution 
the ratio depends on $\sigma_{\varepsilon}$.   
However, the reasoning of \cite{HurowitzCohen2016} implicitly assumes 
that $L$ is long enough, while for our purpose some further refinements 
are crucial.

\sect{Residual bias}
Closing a chain segment of length~$L$ into a ring,
assuming uniform temperature, we get ${\Phi=0}$
due to telescopic cancellations of the terms.
But in the benchmark Sinai model
we have an uncorrelated stochastic field, 
that has dispersion $\sigma_{\varepsilon}$,  
and therefore we get a residual affinity 
${ \Phi \sim \sigma_{\varepsilon} \sqrt{L} }$,
which implies a non-zero residual bias ${ |f| \sim f_L }$, 
where  
\beq \label{eq:res-bias} 
f_L \ \ \equiv \ \ C \dfrac{\sigma_{\varepsilon}}{\sqrt{L}}, 
\ \ \ \ \ \ \  C \approx 0.568
\eeq
The numerical prefactor $C$ is a matter of convention (see below).   
The residual bias $f$ might be either positive or negative
with equal probability, and one expects corresponding 
fluctuations of the spectral density: 
recall that the latter is characterized 
by the $f$-related exponent~$\mu$. 
Those finite size related fluctuations can be ignored
if  ${ f_{1/2} \gg f_L }$, 
which is the implicit assumption in \cite{HurowitzCohen2016}. 
More generally it is  natural to expect ${ f_c \sim f_L }$ 
for a small ring. This expectation is confirmed by \Fig{f4}a,
where the average of $f_c (\sigma_{\varepsilon})$ 
for rings with two different $L$-s is plotted.
The prefector $C$ in the definition \Eq{eq:res-bias} of $f_L$ 
has been determined from this figure.  
The conclusion can be summarized by the formula 
\beq \label{eq:fc-small-sigma}
f_c =  \text{max} \{ f_L, \ f_{1/2} \} 
\sim \text{max} \left\{ \sqrt{\frac{f_{1/2}}{L}} , \ f_{1/2} \right\} 
\ \ \ \
\eeq
From the second equality it should be obvious 
that the crossover takes place at ${ f_{1/2} \sim 1/L }$.  
We emphasize again that for a finite-size ring $f_c$ 
refers to the delocalization transition.

\sect{Irradiated ring}
Turning back to treat our irradiated ring,
the $f_{1/2}$ in \Eq{eq:fc-small-sigma} 
is replaced by an estimate that is based on \Eq{eq:fc}.
However, we have to remember that 
the stochastic field  $\mathcal{E}_n$ does not have Gaussian distribution.
Specifically, in our numerics, 
it reflects the log-normal distribution of the couplings.  
Thus the ratio between  $f_c[\text{delocalization}]$ 
and the $f_c[\text{sliding}]$ estimate of \Eq{eq:fc} 
becomes $\sigma$ dependent. For the ${\sigma=3}$ 
data we used $1/3$ in \Fig{f3}a, and consistently 
the same ratio in \Fig{f3}b.

\sect{Spontaneous delocalization}
Considering a passive particle on a ring without irradiation,
and assuming uniform temperature, we get ${\Phi=0}$
due to telescopic cancellations of the terms. 
But for ${\nu \ne 0}$ we expect a residual bias $f(\nu)$. 
The statistical properties of $f(\nu)$ has been studied in \cite{Hurowitz2013}. 
As in the benchmark Sinai model ${|f(\nu)| \sim f_L}$. 
\Fig{f4}b illustrates this dependence for a few realizations. 
Indeed on the average we witness agreement with \Eq{eq:res-bias}. 
We also plot there (solid line) the expected $f_c$ for a very long ring.
But we are dealing with a finite-$L$ ring, 
and indeed we see that the average $f_c(\nu)$ is in agreement with \Eq{eq:fc-small-sigma}. 
Based on \Eq{fcmax}, it follows that in the vicinity of ${\ln(\nu) \sim 0 }$ 
we should be able to observe the emergence of Sinai physics, i.e. localization, 
provided 
\beq
L \ \ \gg \ \ \frac{1}{\sigma^2 \sigma_v^2}  
\eeq 
Along the tails we observe as expected ${ |f_c(\nu)| \sim f_L(\nu) }$, 
i.e. multiple intersections of the $f_L(\nu)$ curve with the $f_c(\nu)$ curve.  
Thus we have, for either very weak or very strong irradiation,  
a sequence of spontaneous localization-delocalization transitions.        
As ${L \rightarrow \infty}$ the central Sinai region expands.

\section{Discussion}    
\label{sec:discussion}

Testing Sinai physics for a passive particle is possibly rather straightforward. The setup that we have considered in  \Eq{eq:transition-rates} can be literally realized as a semiconductor device. The driving source is like noise that induces extra transition between impurities. 

The analogous discussion for an active particle system is somewhat motivated by experiments with Janus particle that are immersed in a solution. The propulsion can be controlled by an irradiation source. 
But it is not obvious whether the naive setup really leads to the emergence of Sinai physics.  
The model that has been introduced in \cite{Shapira2018}, and the subsequent non-Brownian version that has been considered in \cite{dorkafri2019disorderactive,doussal2020runtumble} assume a static disordered potential $V(x)$ that induces a {\em quenched} velocity field ${v(x) \propto -V'(x)}$. This seems to be a non-physical assumption for a particle in a fluid. The question arises whether it is really feasible to perform an experiment whose objective is to witness induced sub-diffusion and a subsequent sliding transition. 

Our starting point is such that for ${\nu=0}$ we would like to witness normal diffusion. Accordingly we refer to the Brownian version of \Sec{sec:active-brownian}.  If we have in mind Janus particles, it is reasonable to assume that we can induce a {\em uniform} bias. For a charged particle it might be an electric potential ${V(x) \propto -x}$. This 
induces a {\em nonuniform} unidirectional stochastic field ${\mathcal{E}(x)=-\beta(x)V'(x)}$, with $\beta(x)$ that is given, say, by \Eq{eq:beta-x-active}. The $x$~dependence is not due $V(x)$ but due to, say, the disordered irradiation. So the remaining issue is how to get control over the average bias~$f$. In particular how do we achieve ${f=0}$. For that we need an additional mechanism that can counter-balance the average value of $\mathcal{E}(x)$. Apparently the simplest possibility is to consider a Janus particles in a {\em flowing} solution. 

In the setup described above the quench disorder reflects the irradiation pattern, and the bias field is due to 
the net effect of having both external field and flow of the embedding fluid. Such setup allows to test the emergence of Sinai physics for an active Brownian particle.

On the theory side it was important to emphasize a few conceptual subtleties. First of all it should be clear that Sinai model concerns rigorously passive particles. Its application to active particle is not self obvious. We have elaborated several perspectives (heuristic, NESS-based, spectral-based) that support  the emergence of Sinai physics for such non-equilibrium system.          

We have provided estimates for the critical field $f_c(\nu)$. In practice it is \Eq{eq:fc} with $\var{(\beta_n)}$ of \Eq{eq:beta-x-active}. We have tested our estimates numerically as in \Fig{f3}. In order to be able to witness Sinai physics this threshold value should be non-negligible. Specifically, we have clarified how $f_c$ is affected by a finite size effect, and consequently provided the conditions that are required in order to avoid spontaneous delocalization due to a residual bias that is induced by the irradiation source.

\appendix


\section{The formula for the drift velocity}
\label{sec:drift-derrida}

The derivation of the formula for the drift velocity in the continuum limit version of Sinai model is rather simple. One notice that \Eq{eq:fp} is in essence a continuity equation for the current
\beq
I \ &=& \ -D(x)\dfrac{\partial \rho}{\partial x} + v(x) \rho \\
\ &=& \ -D(x) e^{-U(x)} \frac{d}{dx} \left[ e^{U(x)} \rho(x) \right]
\eeq 
The steady state is obtained by solving $I(x) = I = const$, 
where $I$ is yet unknown.
One obtains  
\begin{align} \label{eq:rho-steady-state-w-I}
\rho(x) \ \  = \ \ \left[ C - I \int_0^{x} \frac{ e^{U(x')} }{D(x')} dx' \right] e^{-U(x)} 
\end{align}
where the integration constant $C$ is determined 
by the periodic boundary conditions ${\rho_0(0) = \rho_0(L)}$,
namely,
\beq
C \ \ = \ \ \frac{I}{1-e^{U(L)}} \int_0^{L} \frac{ e^{U(x')} }{D(x')} dx'
\eeq 
At this point it is convenient to make the substitution ${U(x):= [-fx + U(x)]}$, 
such that $U(x)$ is an $L$-periodic function.
Consequently we get 
\beq \label{eq:rho-ss}
\rho(x)  = \frac{I}{1{-}e^{-fL}} 
\int_0^{L} \frac{dr}{D(x{+}r)} e^{ U(x{+}r)-U(x) -fr}  
\eeq
From the normalization condition we get 
the expression for the current:
\beq
I = 
\left( {1{-} e^{fL}} \right)
\left[ \int_0^L \!\!\! \int_0^L dx dr \dfrac{1}{D(x{+}r)} e^{U(x{+}r)-U(x) - fr}  \right]^{-1}
\ \ \ 
\eeq
Replacing the integral over $x$ by an ensemble average,
and using ${I = (1/L)v}$, we get \Eq{vdriftc}.

\section{Non-Brownian active particle}
\label{sec:active-non}

The passive particle that is described by the Fokker Plank Equation \Eq{eq:fp} 
is a Brownian particle whose motion can be described by a Langevin equation    
${ \dot{x} = v(x) + \text{NoiseTerm} }$. 
Without the noise ${D(x)=0}$, which is like having a zero temperature bath.  
We term this noise-less case as the {\em non-Brownian limit}. 
References \cite{dorkafri2019disorderactive,doussal2020runtumble} have considered this limit
for an active particle, aka ``run-and-tumble particle" scenario. 
We summarize their results in the perspective of our presentation.

The motion is described by the Langevin equation 
\beq
\dot{x} \ = \ v(x) \pm \nu 
\eeq
where $\nu$ is the propulsion velocity, and $v(x)$ is the velocity field. 
The $\pm$ sign switches randomly with rate $\alpha$. 
Note that in Ref\cite{dorkafri2019disorderactive} the switching rate is $\alpha/2$, 
while a different notation $\gamma$ is used in \cite{doussal2020runtumble}.
In the tight binding model we use the notation $w^{\alpha}_n$, 
and for uniform irradiation we assume ${w^{\alpha}_n = \const = \alpha }$. 

In the absence of propulsion (${\nu=0}$) we have no diffusion: 
we just have a transient drift due to the local $v(x)$, 
which looks like having zero motion on a coarse grained scale.  
Turning on the propulsion and setting ${v(x)=0}$ 
the coarse grained dynamics of the probability density $\rho(x)$ 
becomes diffusive with 
\beq \label{e31}
D^{\text{eff}} \ = \ \frac{\nu^2}{2\alpha} 
\eeq
This effective diffusion is due to the interplay 
between the propulsion (motion with velocity $\pm \nu$)  
and the random changes in the orientation of the particle (with rate $\alpha$).   
In the presence of ${ v(x) }$ one obtains a quasi-canonical NESS \cite{dorkafri2019disorderactive}:
\beq
\rho(x) \ \propto \  \frac{\nu}{\nu^2-v(x)^2} \, \exp\left[- U(x)\right]
\eeq  
Here $U(x)$ is the effective stochastic potential 
that is associated with with an effective stochastic field   
\beq \label{eEeff}
\mathcal{E}(x)^{\text{eff}} \ = \ \frac{2\alpha v(x)}{\nu^2-v(x)^2}
\eeq
In leading order with respect to ${v(x)}$, 
the result is ${ \mathcal{E}(x) = v(x)/D_0 }$.
We have ${v(x) \propto -V'(x) }$ and therefore 
${U(x) \propto V(x) }$ is bounded, 
which implies that Sinai physics does not emerge. 
But if we go beyond leading order, 
and take the appearance of $v(x)$ in the denominator of \Eq{eEeff} into account, 
then the telescopic correlations are broken, 
$U(x)$ becomes unbounded, and Sinai physics emerges.     
An optional way to break the telescopic correlations 
is to assume randomness in $\alpha$ or in~$\nu$.

\section{Diffusion of active particles}
\label{sec:diffusion-active}

Here we write again the rates of \Sec{sec:active-brownian} with simplified notation
\beq \label{wBrown}
w_{(n{+}1) \uparrow, n \uparrow} &=& D_n +  \frac{v_n}{2} + \nu  \\
w_{n \uparrow, (n{+}1) \uparrow} &=&   D_n -  \frac{v_n}{2} \\
w_{(n{+}1) \downarrow, n \downarrow} &=&  D_n +  \frac{v_n}{2}  \\
w_{n \downarrow, (n{+}1) \downarrow} &=&   D_n -  \frac{v_n}{2} + \nu  \\
w_{n \downarrow, n \uparrow} \ \ \ \ \ &=& \alpha \\
w_{n \uparrow, n \downarrow} \ \ \ \ \ &=& \alpha
\eeq
In the absence of disorder we set $D_n=D_0$ and $v_n=v_0$. 
The calculation of the effective diffusion coefficient,  
is done with the help of Bloch theorem.
In the momentum basis, 
the rate matrix is block diagonal in~$k$.
The blocks are
\begin{align}\label{eq:w-mat-k}
\bm{W}^{(k)} = 
\left(\begin{matrix}
A_{+}^{(k)} - A_{+}^{(0)} - \alpha & \alpha \\
\alpha & A_{-}^{(k)} - A_{-}^{(0)} - \alpha  
\end{matrix} \right)
\end{align}
where 
\begin{align*}
A_{\pm}^{(k)} =  2 D_0 \cos{(k)} - i v_0 \sin{(k)  + \nu e^{\mp i k}}
\end{align*}
The eigenvalues are denoted $-\lambda_{k,\pm}$.  
The drift velocity and diffusion coefficient are found 
by expanding ${ \lambda_{k,+} }$ as in \Eq{eSpectrum}.
The drift velocity comes out $v_0$ as expected, 
and for the diffusion coefficient we get
\beq \label{dBrown}
D^{\text{eff}} \ = \ D_0 + \frac{\nu}{2} + \frac{\nu^2}{2\alpha}
\eeq 
In terms of the parameters that are used in the main text, we get \Eq{eDeff}.

For completeness we write the explicit expression for the eigenvalues. 
In the absence of bias this expression is rather simple:   
\begin{align}\label{eq:lambdak}
\lambda_{k,\pm} = -\left[ c \pm \sqrt{\alpha^2-a^2}\, \right]
\end{align}
with
\begin{align*}
\alpha &= w^{\alpha} \\
a &= w^{\nu} \sin(k)  \\
c &= (2 w^{\beta} + w^{\nu}) \cos(k) 
- \left[ w^{\alpha} + (2 w^{\beta} + w^{\nu}) \right]
\end{align*}
For ${w^{\nu} > w^{\alpha}}$ some of the eigenvalues become complex, 
but the small~$k$ eigenvalues at the vicinity of ${\lambda=0}$ 
are always real. The latter statement assumes a very long ring.  
For a finite ring of length~$L$ such eigenvalues (with ${ k \sim 1/L }$) exist 
provided ${ w^{\nu} }$ is smaller compared with ${ w^{\alpha} L }$.

\sect{Band mixing}
\rmrk{The essence of Sinai physics is that weak bias 
is not able to make the spectrum complex, 
as opposed to the non disordered result \Eq{eSpectrum}. 
Instead, $f$  has to exceed a threshold to make the spectrum complex. 
It is important to remember that we care here about the 
spectrum at the vicinity of ${\lambda=0}$. 
The above perspective applies also for an active particle. 
Namely, also for an {\em active particle in a disordered environment} 
the spectrum at the vicinity of ${\lambda=0}$ remains real if $f$ is small enough.
This observation is based on the following reasoning. 
The diagonalization of ${\bm{W} = \bm{H} + \bm{A} }$ 
can be treated within the framework of perturbation theory 
as explained in Section~V of \cite{Shapira2018}.
Here $\bm{H}$ is a real symmetric matrix, 
while $\bm{A}$ is a real anti-symmetric matrix.
It is well known (usually in the context of PT-symmetric non-hermitian Hamiltonians) 
that in order to get complex spectrum the elements of $\bm{A}$ should be 
large enough in the sense of perturbation theory (the coupling should be larger 
than the ``energy-difference" of the coupled states),   
else the spectrum remains {\em real}.    
In the present context $\bm{H}$ is composed of two uncoupled $\pm$ blocks, 
while ${\bm{A}= \bm{A}_{\perp} + \bm{A}_{\parallel} }$ 
is composed of propulsion-related-perturbation and bias-related-perturbation, respectively.  
The former induces inter-band coupling, 
while the latter induces intra-band couplings.
Without $\bm{A}_{\perp}$ we have the standard delocalization scenario:
if the intra-band non-hermitian perturbation is strong enough 
one observes Sinai-type delocalization (complex eigenvalues).    
On the other hand, without $\bm{A}_{\parallel}$, 
we have only inter-band couplings that are likely to induce complexity 
only in the band-overlap region. This argument assumes that the propulsion 
is not too extreme, such that we stay within the Sinai regime. 
In other words: we can establish the emergence of Sinai physics for 
active particles, but we cannot exclude the existence of additional regimes.}

\section{NESS for non-Brownian active particle} 
\label{sec:ness-active}

The NESS can be easily found for a unidirectional stochastic motion. The propulsion rate~$\nu$ is modulated by the velocity field of the disordered bath, and the counter-propulsion rates are neglected. Instead of the rates as written in \App{sec:active-brownian} we use the following non-Brownian version:
\beq
w_{(n{+}1) \uparrow, n \uparrow} &=&  \nu + v_n \ \equiv \ w_n^{+} \\
w_{n \uparrow, (n{+}1) \uparrow} &=&   0 \\
w_{(n{+}1) \downarrow, n \downarrow} &=&  0  \\
w_{n \downarrow, (n{+}1) \downarrow} &=&   \nu - v_n \ \equiv \ w_n^{-} \\
w_{n \downarrow, n \uparrow} \ \ \ \ \ &=& \alpha \\
w_{n \uparrow, n \downarrow} \ \ \ \ \ &=& \alpha
\eeq
One can verify that for a non-disordered system the diffusion coefficient 
is the same, namely \Eq{dBrown}, but without the first term. 

\rmrk{The NESS is determined by the equation ${\bm{W}\bm{p}=0}$. 
In the Sinai regime the drift velocity vanishes. 
Formally this statement is exact only 
in a statistical sense (for an infinite chain).
Consequently, for a very long ring we expect 
an exponentially small NESS current.  
Below we find the zero-current solution. 
It can be regarded as an approximation for the exact NESS.
Clearly we are using here a self-consistent approximation scheme:  
outside of the Sinai regime this scheme fails.  
The extreme illustration for the failure of this procedure 
is the case of a non-disordered ring 
(in the continuum limit it is \Eq{eq:fp} with constant~$D$ and~$v$). 
The zero current solution for non-zero~$v$ 
is exponentially diverging in one direction, 
and cannot be matched if one tries to impose 
periodic boundary conditions. The true NESS in the latter 
case does not feature an exponentially small current.}

The equation for the zero-current NESS for non-Brownian active particle is 
\beq
p_{n{+}1,\downarrow} \ \ = \ \ \left( \frac{w_n^{+}}{w_n^{-}} \right) \ p_{n,\uparrow}
\eeq
From the Kirchhoff equation at node ${ (n{+}1,\uparrow) }$ we deduce that 
\beq
\frac{p_{n{+}1,\uparrow}}{p_{n,\uparrow}} \ \ = \ \ 
\left[\frac{\alpha+w_{n}^{-}}{\alpha+w_{n{+}1}^{+}} \right] \left( \frac{w_n^{+}}{w_n^{-}} \right)
\eeq 
From here it follows that 
\beq
\frac{p_{n{+}1}}{p_{n}} 
\ = \ 
\frac
{\left[ 1 + \frac{\alpha+w_{n}^{-}}{\alpha+w_{n{+}1}^{+}} \right]}  
{\left[ 1 + \frac{\alpha+w_{n}^{+}}{\alpha+w_{n{-}1}^{-}} \right]}
\left(\frac{w_n^{+}}{w_n^{-}} \right)
\eeq
The effective stochastic field is found via \Eq{e7}.

\clearpage
 
\sect{Acknowledgment}
This research was supported by the Israel Science Foundation (Grant No.283/18).  

\ \\ \ \\ 



\begin{thebibliography}{10}

\bibitem{Sinai1983}
Sinai Y~G 1982 {\em Theory of Probability \& Its Applications\/} {\bf 27}  256--268 
\hrefl{[link]}{https://doi.org/10.1137/1127028}

\bibitem{Derrida1983}
Derrida B 1983 {\em Journal of Statistical Physics\/} {\bf 31} 433--450 ISSN 1572-9613 
\hrefl{[link]}{https://doi.org/10.1007/BF01019492}

\bibitem{BOUCHAUD1990}
Bouchaud J~P, Comtet A, Georges A and Le~Doussal P 1990 {\em Annals of Physics\/} {\bf 201} 285--341
\hrefl{[link]}{http://www.sciencedirect.com/science/article/pii/000349169090043N}

\bibitem{BOUCHAUD1990a}
Bouchaud J~P and Georges A 1990 {\em Physics Reports\/} {\bf 195} 127 -- 293 
\hrefl{[link]}{http://www.sciencedirect.com/science/article/pii/037015739090099N}

\bibitem{Creep}
Le~Doussal P and Vinokur V~M 1995 {\em Physica C: Superconductivity\/} {\bf 254} 63--68
\hrefl{[link]}{https://www.sciencedirect.com/science/article/abs/pii/0921453495005455}

\bibitem{Hatano1996}
Hatano N and Nelson D~R 1996 {\em Phys. Rev. Lett.\/} {\bf 77}(3) 570--573
\hrefl{[link]}{https://link.aps.org/doi/10.1103/PhysRevLett.77.570}

\bibitem{Hatano1997}
Hatano N and Nelson D~R 1997 {\em Phys. Rev. B\/} {\bf 56}(14) 8651--8673
\hrefl{[link]}{https://link.aps.org/doi/10.1103/PhysRevB.56.8651}

\bibitem{Shnerb1998}
Shnerb N~M and Nelson D~R 1998 {\em Physical review letters\/} {\bf 80} 5172
\hrefl{[link]}{http://journals.aps.org/prl/pdf/10.1103/PhysRevLett.80.5172}

\bibitem{Feinberg1999}
Feinberg J and Zee A 1999 {\em Phys. Rev. E\/} {\bf 59}(6) 6433--6443
\hrefl{[link]}{https://link.aps.org/doi/10.1103/PhysRevE.59.6433}

\bibitem{KAFRI2004}
Kafri Y, Lubensky D~K and Nelson D~R 2004 {\em Biophysical Journal\/} {\bf 86} 3373 -- 3391 
\hrefl{[link]}{http://www.sciencedirect.com/science/article/pii/S0006349504743854}

\bibitem{KAFRI2005}
Kafri Y, Lubensky D~K and Nelson D~R 2005 {\em Phys. Rev. E\/} {\bf 71}(4) 041906 
\hrefl{[link]}{https://link.aps.org/doi/10.1103/PhysRevE.71.041906}

\bibitem{HurowitzCohen2016}
Hurowitz D and Cohen D 2016 {\em Scientific reports\/} {\bf 6}
  \hrefl{[link]}{http://www.nature.com/articles/srep22735}

\bibitem{HurowitzCohen2016a}
Hurowitz D and Cohen D 2016 {\em Phys. Rev. E\/} {\bf 93}(6) 062143
\hrefl{[link]}{https://link.aps.org/doi/10.1103/PhysRevE.93.062143}

\bibitem{stein1995broken}
Stein D~L and Newman C~M 1995 {\em Physical Review E\/} {\bf 51} 5228
\hrefl{[link]}{https://journals.aps.org/pre/abstract/10.1103/PhysRevE.51.5228}

\bibitem{reichhardt2014active}
Reichhardt C and Reichhardt C~O 2014 {\em Physical Review E\/} {\bf 90} 012701
\hrefl{[link]}{https://journals.aps.org/pre/abstract/10.1103/PhysRevE.90.012701}

\bibitem{reichhardt2018avalanche}
Reichhardt C~O and Reichhardt C 2018 {\em New Journal of Physics\/} {\bf 20} 025002
  \hrefl{[link]}{https://iopscience.iop.org/article/10.1088/1367-2630/aaa392/meta}

\bibitem{reichhardt2017negative}
Reichhardt C and Reichhardt C~J~O 2017 {\em Journal of Physics: Condensed
  Matter\/} {\bf 30} 015404
  \hrefl{[link]}{https://iopscience.iop.org/article/10.1088/1361-648X/aa9c5f/meta}

\bibitem{reichhardt2021clogging}
Reichhardt C and Reichhardt C~J~O 2021 {\em Physical Review E\/} {\bf 103} 062603
\hrefl{[link]}{https://journals.aps.org/pre/abstract/10.1103/PhysRevE.103.062603}

\bibitem{bertrand2018optimized}
Bertrand T, Zhao Y, B{\'e}nichou O, Tailleur J and Voituriez R 2018 {\em Physical Review Letters\/} {\bf 120} 198103
\hrefl{[link]}{https://journals.aps.org/prl/abstract/10.1103/PhysRevLett.120.198103}

\bibitem{Walther2013}
Walther A and M{\"u}ller A~H~E 2013 {\em Chemical reviews\/} {\bf 113} 5194--5261 
\hrefl{[link]}{https://pubs.acs.org/doi/abs/10.1021/cr300089t}

\bibitem{Wheat2010}
Wheat P~M, Marine N~A, Moran J~L and Posner J~D 2010 {\em Langmuir\/} {\bf 26}
  13052--13055 \hrefl{[link]}{https://pubs.acs.org/doi/abs/10.1021/la102218w}

\bibitem{Menzel2015}
Menzel A~M 2015 {\em Physics reports\/} {\bf 554} 1--45
\hrefl{[link]}{https://www.sciencedirect.com/science/article/pii/S0370157314003871}

\bibitem{maggi2015}
Maggi C, Simmchen J, Saglimbeni F, Katuri J, Dipalo M, Angelis F~D, Sanchez S and Leonardo R~D 2015 {\em Small\/} {\bf 12} 446--451
\hrefl{[link]}{https://onlinelibrary.wiley.com/doi/abs/10.1002/smll.201502391}

\bibitem{Hurowitz2013}
Hurowitz D, Rahav S and Cohen D 2013 {\em Physical Review E\/} {\bf 88} 062141
\hrefl{[link]}{https://doi.org/10.1103/PhysRevE.88.062141}

\bibitem{Shapira2018}
Shapira D, Meidan D and Cohen D 2018 {\em Phys. Rev. E\/} {\bf 98}(1) 012107
\hrefl{[link]}{https://link.aps.org/doi/10.1103/PhysRevE.98.012107}

\bibitem{dorkafri2019disorderactive}
Dor Y~B, Woillez E, Kafri Y, Kardar M and Solon A~P 2019 {\em Physical Review  E\/} {\bf 100} 052610
\hrefl{[link]}{https://journals.aps.org/pre/abstract/10.1103/PhysRevE.100.052610}

\bibitem{doussal2020runtumble}
Le~Doussal P, Majumdar S~N and Schehr G 2020 {\em EPL (Europhysics Letters)\/} {\bf 130} 40002
\hrefl{[link]}{https://iopscience.iop.org/article/10.1209/0295-5075/130/40002/pdf}

\bibitem{hanggi2009artificial}
H{\"a}nggi P and Marchesoni F 2009 {\em Reviews of Modern Physics\/} {\bf 81}  387
\hrefl{[link]}{https://journals.aps.org/rmp/abstract/10.1103/RevModPhys.81.387}

\end{thebibliography}


\end{document}